\documentclass[aps,prb,reprint,superscriptaddress,nofootinbib,floatfix]{revtex4-2}
\usepackage{amssymb}
\usepackage{physics}
\usepackage{graphicx}
\usepackage{amsmath}
\usepackage{amsthm}
\usepackage{amsfonts}
\usepackage[T1]{fontenc}
\usepackage{array}
\usepackage{multirow}
\usepackage{color}
\usepackage{esint}
\usepackage{bm}
\usepackage{color}
\usepackage{bbm}
\usepackage{hyperref}
\usepackage{babel}

\newcommand{\modelname}{$t_1$-$t_2$}

\usepackage{hyperref}
\usepackage[normalem]{ulem}
\hypersetup
{	colorlinks,%
	citecolor=green,%
	linkcolor=red,%
	urlcolor=blue%
}
\allowdisplaybreaks[4]
\begin{document}

\global\long\def\id{\mathbbm{1}}
\global\long\def\ui{\mathbbm{i}}
\global\long\def\ud{\mathrm{d}}

\title{Statistics of noninteracting many-body fermionic states: The question of a many-body mobility edge}

\author{Ke Huang}
\affiliation{Department of Physics, City University of Hong Kong, Kowloon, Hong Kong SAR, China}
\author{DinhDuy Vu}
\affiliation{Condensed Matter Theory Center and Joint Quantum Institute, University of Maryland, College Park, Maryland 20742, USA}
\author{Sankar Das Sarma}
\affiliation{Condensed Matter Theory Center and Joint Quantum Institute, University of Maryland, College Park, Maryland 20742, USA}
\author{Xiao Li}
\email{xiao.li@cityu.edu.hk}
\affiliation{Department of Physics, City University of Hong Kong, Kowloon, Hong Kong SAR, China}
\date{\today}

\begin{abstract}
In this work, we study the statistics of a generic noninteracting many-body fermionic system whose single-particle counterpart has a single-particle mobility edge (SPME). 
We first prove that the spectrum and the extensive conserved quantities follow the multivariate normal distribution with a vanishing standard deviation $\sim O(1/\sqrt L)$ in the thermodynamic limit, regardless of SPME. 
Consequently, the theorem rules out an infinite-temperature or high-temperature many-body mobility edge (MBME) for generic noninteracting fermionic systems. 
Further, we also prove that the spectrum of a fermionic many-body system with short-range interactions is qualitatively similar to that of a noninteracting many-body system up to the third-order moment.
These results partially explain why neither short-range~\cite{Huang2023} nor long-range interacting systems exhibit an infinite-temperature MBME. 
\end{abstract}

\maketitle
\section{Introduction}
Strong disorder is a generic mechanism to create localized phases in both single-particle and many-body systems. 
In single-particle systems, strong disorder renders an extended system localized, a phenomenon known as Anderson localization~\cite{Anderson1958}. 
Similarly, a many-particle system with strong disorder can become many-body localized (MBL) and avoid thermalization~\cite{Nandkishore2015_Review,Altman2015_Review,Imbrie2016,Imbre2016,Abanin2019_RMP,Gopalakrishnan2020_Review}. 
Roughly speaking, the extended and localized phases in the single-particle system ``correspond'' to the ergodic and the nonergodic phases in the many-body system. 
Beyond the extended and the localized phases, the single-particle system can also be in an intermediate phase with coexisting extended and localized eigenstates. 
In the intermediate phase, both the extended and localized states occupy a finite fraction of the whole spectrum. 
Furthermore, the two parts of the spectrum are energetically separated by the single-particle mobility edge (SPME)~\cite{DasSarma1986_PRL,DasSarma1988_PRL,Thouless1988_PRL,Biddle2010_PRL,Biddle2011_PRB,Ganeshan2015_PRL,Li2017_PRB,Li2020_PRB,Wang2020}.

The many-body counterpart of the SPME is suggested to be the many-body mobility edge (MBME), which may imply different concepts depending on the context and the model. 
Many theoretical works on MBME~\cite{Gornyi2005,Basko2006,Kjall2014,Luitz2015,Mondragon2015,LiXiaopeng2015,Devakul2015,Nag2017,Villalonga2018,Wei2019,Chanda2020,wei2020,Yousefjani2023} consider the critical temperature at which the system experiences the thermal-MBL transition. 
From a dynamical perspective, this amounts to determining whether the randomly chosen initial states at different energies relax similarly. 
Therefore, we call it the temperature-dependent MBME. 
This type of MBME has been numerically and experimentally~\cite{Guo2020} found even in systems without SPME. 
However, this subject is controversial, as these results were mostly obtained in small systems, and some arguments~\cite{Roeck2016} suggest that the temperature-dependent MBME may not exist in the thermodynamic limit.
The extreme case is the infinite-temperature MBME, which is equivalent to examining whether the randomly chosen initial states (without controlling their energies~\cite{Kohlert2019_PRL}) have similar dynamics. 
Hence, an infinite-temperature MBME entails finite fractions of both extended and localized states, so that the possibility of (randomly) choosing either is nonzero. 
In the presence of an SPME, as there are finite fractions of both extended and localized single-particle orbitals, an intuitive conjecture is that there are also finite fractions of different types of many-body states. 
Hence, in this work, we adopt the convention that a system is in the many-body intermediate phase if it has an infinite-temperature MBME. 
The ``intermediate'' here alludes to a possible dynamical phase in between purely thermal and purely MBL at the infinite temperature~\cite{Gornyi2005,Basko2006,Kjall2014,Luitz2015,Mondragon2015,LiXiaopeng2015,Devakul2015,Nag2017,Villalonga2018,Wei2019,Chanda2020,wei2020,Yousefjani2023}~\footnote{Note that in the single-particle systems, a finite-temperature ME always implies an infinite-temperature ME because the density of state (DOS) is finite almost everywhere.}. 



Our previous work~\cite{Huang2023} specifically studied the \modelname\ model (which has an SPME) with short-range (SR) interactions, in which the SPME does not induce an infinite-temperature MBME, though the signature of MBME at low temperatures has been found in small systems. 
In the current work, we first analyze the \modelname\ model with long-range (LR) interactions, which turns out to be qualitatively similar to the SR interaction case for intermediate interacting strength and has no infinite-temperature MBME as well. 
Thus, we go back to the noninteracting model to see whether the ineffectiveness of the SPME in inducing an MBME is a consequence of interaction at all, or whether it arises from some intrinsic single-particle properties.  
We do this by studying the noninteracting many-body fermionic system. 

Note that a recent work~\cite{lefevre2022} introduced a numerical algorithm to efficiently obtain the exact density of states (DOS) in finite-size noninteracting systems. 
In this work, however, we focus on the statistical behavior in the thermodynamic limit. 
Though the thermal-MBL transition at a generic finite temperature is a rather difficult problem, we are able to establish at high and infinite temperature that, for noninteracting fermions, the DOS and other extensive quantities commuting with the Hamiltonian form a multivariate normal distribution with parameters determined by the expectation value and the variance of the single-particle system. 
This theorem implies that all specific single-particle structures, including the SPME, are smeared out in the many-body system. 
Moreover, the standard deviation of the many-body energy eigenvalues scales as $O(1/\sqrt L)$, and therefore, the DOS is zero almost everywhere in the many-body system, in contrast to the single-particle case. 
Consequently, the existence of a finite-temperature MBME does not guarantee an infinite-temperature MBME.
Further, the energy-resolved standard deviation of extensive quantities also manifests the $O(1/\sqrt L)$ scaling, which precludes both the infinite- and high-temperature MBME in noninteracting fermionic systems. 
Finally, we analyze the effect of the SR interaction on the noninteracting spectrum. 
Limited to small system sizes, the numerical results suggest that the many-body spectrum of the interacting system has a broader width and a finite skewness signifying an asymmetry. 
However, we demonstrate that these are finite-size effects, as we show that up to the third-order moment, the spectrum of the interacting systems also follows the $O(1/\sqrt L)$ scaling.

The structure of the paper is the following. 
In Section~\ref{Sec:LR}, we analyze the effect of the LR interaction in the \modelname\ model, demonstrating that no finite-temperature MBME emerges in the numerical results. 
In Section~\ref{Sec:Noninteracting}, we present the central result of this work. 
Specifically, we show that due to the central limit theorem (CLT), most details in the single-particle spectrum are washed out when we build the many-body states. 
This result explains why the SPME is ineffective in inducing the infinite-temperature MBME. 
In Section~\ref{Sec:SR}, we analyze how the spectrum of a noninteracting fermionic system is modified by short-range interactions. 
Finally, in Section~\ref{Sec:Discussion}, we provide some more discussions and conclude this paper.

\section{long-range interaction \label{Sec:LR}}
\begin{figure}[!]
\includegraphics[width=\columnwidth]{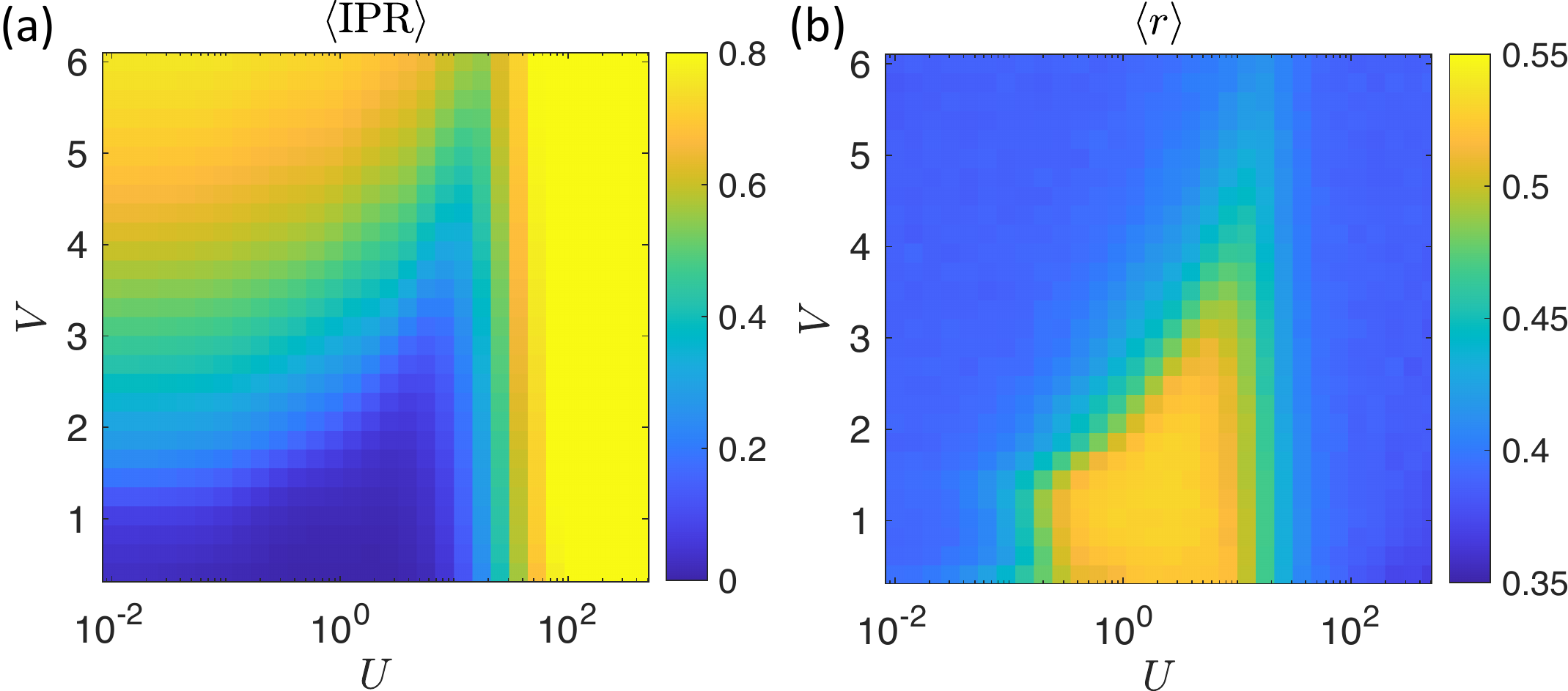}
\caption{\label{Fig:IPR&LS}
(a) IPR and (b) mean gap ratio of the \modelname\ model with long-range interaction. Here, we take $t_2=1/6$, and the system size is $L=16$. The results are averaged over six random phases.
}
\end{figure}

In this section, we consider the \modelname\ model with LR interaction at half-filling, given by
\begin{align}
	H=\sum_j\left(t_1c^\dag_j c_{j+1}+t_2c^\dag_j c_{j+2}+\text{H.c.}\right)\nonumber\\
	+\sum_jV_jn_j+\frac12\sum_{i\neq j}U_{\abs{i-j}}n_in_j,
\end{align}
where $V_j=V\cos(2\pi q j+\phi)$ is the quasiperiodic potential with $q=(\sqrt{5}-1)/2$, and $\phi$ is a random phase. Here, we adopt the periodic boundary condition and study the Coulomb interaction given by
\begin{align}
	U_r=U\left(\frac{L}{\pi}\sin\frac{\pi r}L\right)^{-1}.
\end{align}
Additionally, we set $t_1=1$ as the energy unit and take $t_2=1/6$.
To quantify the localization of the many-body eigenstates, we utilize the many-body inverse participation ratio (MIPR), defined as~\cite{Vu2022_PRL} 
\begin{align}
	\mathcal I=\frac1{1-N/L}\qty[\frac{1}{N}\sum_{i=1}^L\bar{n}_i^2-N/L], 
\end{align}
where $\bar{n}_i$ is the particle number expectation value on site $i$. 
The MIPR describes the density distribution in real space, and we have $\mathcal I\to 0$ for an extended state, while $\mathcal I\to 1$ for a localized state. 
In Fig.~\ref{Fig:IPR&LS}(a), we calculate the MIPR of the \modelname\ model in a system of length $L=16$, which exhibits qualitatively similar behavior as the AA model and the Anderson model with LR interactions~\cite{Vu2022_PRL}. 
Specifically, the \modelname\ model with LR interaction possesses three distinct regimes for different interaction strengths: the single-particle regime $U\ll O(1)$, the many-body regime $U\sim O(1)$, and the Mott regime $U\gg O(1)$. 
Compared with the \modelname\ model with SR interaction~\cite{Huang2023}, the LR interaction manifests no qualitative difference in the single-particle and the many-body regimes. 
However, the LR interaction fundamentally alters the Hilbert space structure in the Mott regime, resulting in localization at infinitesimal disorder strengths.

We also calculate the mean gap ratio, which is another way to characterize the MBL transition. 
The gap ratio is defined by
\begin{align}
	r=\min\left\{\frac{\delta E_i}{\delta E_{i+1}},\frac{\delta E_{i+1}}{\delta E_i}\right\},
\end{align}
where $\delta E_i=E_{i+1}-E_i$ is the energy gap between two adjacent eigenenergies. 
In the thermal phase, the spectrum follows the Gaussian orthogonal ensemble (GOE) whose mean gap ratio $\expval{r}$ is $0.53$, whereas the spectrum follows the Poisson distribution in the MBL phase with $\expval{r}=0.386$. 
As shown in Fig.~\ref{Fig:IPR&LS}(b), the interaction range has no significant effect in the single-particle and the many-body regimes. 
However, the thermal Mott regime for the SR interaction disappears in the presence of the LR interaction. 
This phenomenon is subtly different from the corresponding situation in the AA model and the Anderson model, which are not thermal in the Mott regime for the SR interaction because of the Hilbert-space fragmentation~\cite{DeTomas2019_PRB,Yang2020_PRL,Vu2022_PRL}. 
The similarity between SR and LR cases in the many-body regime suggests that the LR interaction cannot induce a many-body intermediate phase arising from the single-particle intermediate phase.

\section{Noninteracting fermions \label{Sec:Noninteracting}}
In this section we analyze the statistical properties of noninteracting many-body fermionic states. 
We will discuss three questions. 
In Section~\ref{SubSec:A}, we introduce a revised central limit theorem (CLT) for noninteracting fermions. 
In Section~\ref{SubSec:B}, we use this result to argue that the MIPR must be a continuous function of energy, which rules out the high-temperature MBME in a noninteracting fermionic system. 
Finally, in Section~\ref{SubSec:C} we show that the MIPR in a half-filled fermionic system must be an even function of energy~\footnote{We always set the mean energy as zero in this work.}. 

\subsection{A revised central limit theorem for noninteracting fermions \label{SubSec:A}}
The results for the SR and LR interactions in the \modelname\ model hint that the ineffectiveness of the SPME in introducing an MBME in the many-body system may not be caused by the interaction. 
Hence, we take a step back and study the noninteracting many-body system. 
In a noninteracting classical system where there is no indistinguishability, each particle can be regarded as an independent variable. 
Hence, the statistics of an extensive quantity $X_{\text{tot}}=\sum_{i=1}^N X_i$ follows the CLT
\begin{align}
	\frac{X_{\text{tot}}-\mu_xN}{\sqrt{N}\sigma_x}\sim \mathcal{N}(0,1),
\end{align}
where $\mu_x$ and $\sigma_x^2$ are the expectation value and the variance of $X_i$, and $\mathcal{N}(\mu,\sigma^2)$ is the univariate normal distribution with mean $\mu$ and standard deviation $\sigma$.   

However, the CLT does not apply to fermionic systems, as the Pauli exclusion principle implies that particles are not independent variables. 
In fact, building a many-body state in a fermionic system is equivalent to sampling without replacement. 
If a quantity $X$ of orbital $i$ takes value $x_i$ and there are $L$ orbitals, then calculating $X_{\text{tot}}$ of $N$ particles amounts to choose $N$ samples out of $L$ objects. 
Further, we have to deal with the multivariate distribution if there are multiple quantities. 
Hence, we can ask whether there is a revised CLT for a noninteracting many-body fermionic system, where all the variables are not independent. 

To begin with, we consider the case with just two variables. 
Suppose that we have two sets of values $X=\{x_i\}_{i=1}^L$, $Y=\{y_i\}_{i=1}^L$, and $\{X_j\}_{j=1}^N$, $\{Y_j\}_{j=1}^N$ ($N<L$) are the random variables representing the sampling without replacement. 
Because of the permutation symmetry, we know that the probability $P(X_j=x_i)=P(X_1=x_1)=1/L$ and that $P(X_j=x_i|X_k=x_l)=(1-\delta_{il})/(L-1)$ if $j\neq k$. 
As a result, the expectation value of $(X_j-\mu_x)(X_k-\mu_x)$ is 
\begin{widetext}
\begin{align}
	\mathbb{E}\qty[(X_j-\mu_x)(X_k-\mu_x)]
    =&\sum_{i=1}^L(x_i-\mu_x)P(X_j=x_i)\cdot\sum_{l=1}^L (x_l - \mu_x) P(X_k=x_l|X_j=x_i)
	=\frac1{L(L-1)}\sum_{i\neq l}(x_i-\mu_x)(x_l-\mu_x)\nonumber\\
	=&\frac1{L(L-1)}\left[\left(\sum_{i}x_i-\mu_xL\right)^2-\sum_{i}(x_i-\mu_x)^2\right]
	=\frac{\sigma_x^2}{L-1}. 
\end{align}
\end{widetext}
Consequently, we have 
\begin{align}
	\mathbb E(X_{\text{tot}}-\mu_xN)^2=\frac{N(L-N)}{L-1}\sigma_x^2.
\end{align}
Similarly, we have
\begin{align}
	\text{Cov}(X_{\text{tot}},Y_{\text{tot}})=\frac{N(L-N)}{L-1}\text{Cov}(X,Y),
\end{align}
where $\text{Cov}(X,Y)$ is the covariance between $X$ and $Y$.  

\begin{figure}[!]
	\includegraphics[width=\columnwidth]{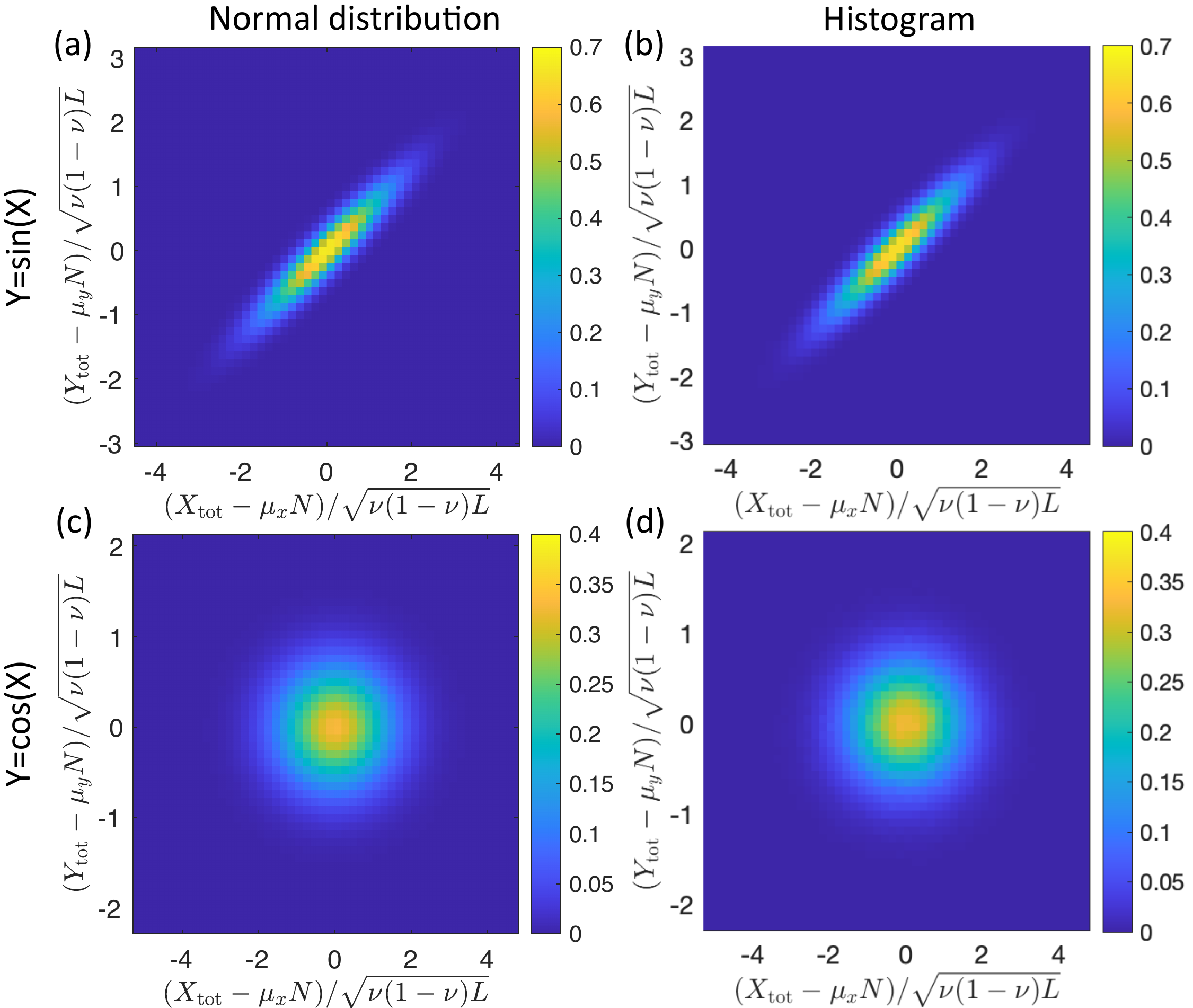}
	\center
	\caption{\label{Fig:CLT}The normal distribution predicted by CLT (left panel) and the histogram of $X_{\text{tot}},Y_{\text{tot}}$ (right panel). Further, we consider the antisymmetric case $Y=\sin(X)$ in the upper panel, and the symmetric case $Y=\cos(X)$ in the lower panel.
	}
\end{figure}

More generally, we prove in Appendix~\ref{SM:CLT} that all the moments agree with those of the multivariate normal distribution in the thermodynamic limit. 
This implies that if $\Sigma$, the covariance matrix between $X$ and $Y$, is positive definite, we have 
\begin{align}\label{Eq:CLT}
	\frac{X_{\text{tot}}-\mu_xN}{\sqrt{\nu(1-\nu)L}},\frac{Y_{\text{tot}}-\mu_yN}{\sqrt{\nu(1-\nu)L}}\sim \mathcal{N}(0,\Sigma),
\end{align}
where $\mathcal{N}(0,\Sigma)$ denotes the multivariate normal distribution, and $\nu=N/L$ is the filling factor. 
Note that the univariate case of the above result has been proven in Ref.~\cite{Ethier2010doctrine}. 
In Fig.~\ref{Fig:CLT}, we present a numerical example of this CLT. 
In the right panel, we set $L=1000$, $N=300$ and take either $Y=\cos(X)$ or $Y=\sin(X)$, which agrees perfectly with the normal distribution $\mathcal N(0,\Sigma)$ shown in the left panel. 
Here, the value of $x_i$ is generated by the normal distribution, and, as we cannot obtain all the product states, we randomly choose $10^6$ product states (with replacement) as the sample.

A crucial implication of Eq.~\eqref{Eq:CLT} is that the conditional expectation of $Y_{\text{tot}}$ depends on $X_{\text{tot}}$ linearly.
In particular, we have
\begin{align}
	\mathbb{E}(Y_\text{tot}|X_\text{tot}=x)=
	\frac{\text{Cov}(X,Y)}{\sigma_x^2}(x-\mu_x N)+\mu_y N, \label{Eq:Linear}
\end{align}
whose generic behavior is shown in Fig.~\ref{Fig:CLT}(a). 
Further, if $Y=f(X)$ and $f$ is an even function, $\text{Cov}(X,Y)=0$ and therefore $X_{\text{tot}},Y_{\text{tot}}$ are independent. 
Hence, the conditional expectation is constant. 
This special case is shown in Fig.~\ref{Fig:CLT}(c), where the highlighted region has a circular shape. 

Furthermore, Eq.~\eqref{Eq:CLT} also suggests the equivalence of the microcanonical ensemble and the canonical ensemble at high temperatures. 
We first consider the infinite temperature case, where the system is controlled by the DOS of the many-body spectrum. 
We take $X_{\text{tot}}$ to be the total energy, and thus we have the analytic expression of the DOS, given by 
\begin{align}
	g(\varepsilon)\propto\exp[-\frac N2\frac{(\varepsilon-\mu_\varepsilon)^2}{\sigma_\varepsilon^2(1-\nu)}],
\end{align}
where $\varepsilon=E/N$ is the energy per particle. 
Note that the energy per particle is equivalent to the energy density up to a linear transformation, and the latter has been used extensively in the studies of MBME~\cite{Gornyi2005,Basko2006,Kjall2014,Luitz2015,Mondragon2015,LiXiaopeng2015,Devakul2015,Nag2017,Villalonga2018,Wei2019,Chanda2020,wei2020,Yousefjani2023}. 
Therefore, in the thermodynamic limit, almost all the states reside in a small energy interval centered at $\mu_\varepsilon$ of width $O(1/\sqrt{N})$, which is essentially the microcanonical ensemble. 

At any finite temperature, the density matrix of the canonical ensemble is given by
\begin{align}
	\rho=\frac{e^{-\beta H}}{\trace[e^{-\beta H}]}\propto\int\dd \varepsilon\dyad{\varepsilon}e^{-\beta N\varepsilon}g(\varepsilon),
\end{align}
where $\beta$ is the inverse temperature. 
Thus, in the high-temperature limit ($\beta\ll1$), we obtain
 \begin{align}
	e^{-\beta N\varepsilon}g(\varepsilon)\propto\exp[-\frac N2\frac{(\varepsilon-\tilde\mu_\varepsilon)^2}{\sigma_\varepsilon^2(1-\nu)}], \label{Eq:Temp-Shift}
\end{align}
where $\tilde\mu_\varepsilon=\mu_\varepsilon-\beta(1-\nu)\sigma_\varepsilon^2$. Hence, finite temperature just means shifting the center of the energy interval from $\mu_\varepsilon$ to $\tilde\mu_\varepsilon$, and the canonical ensemble is still equivalent to the microcanonical ensemble in this case.

\begin{figure*}[t]
	\includegraphics[width=\textwidth]{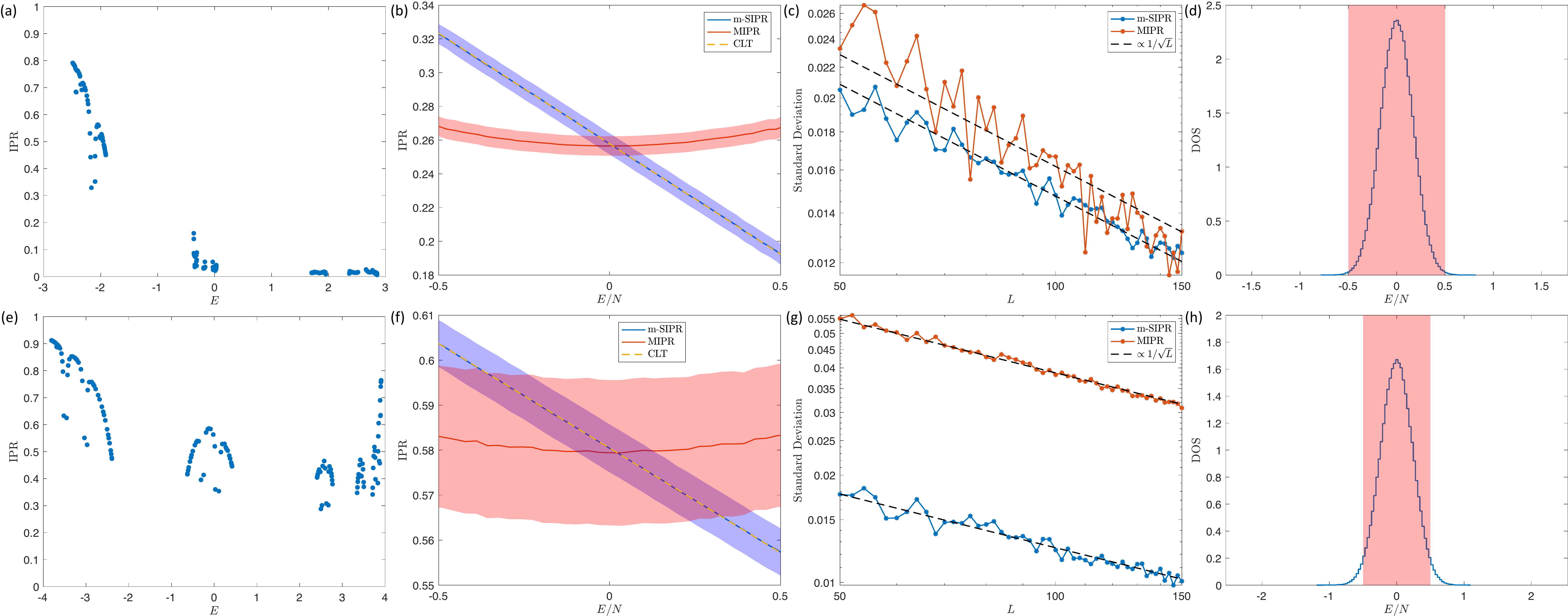}
	\center
	\caption{\label{Fig:IPR}(a-d): \modelname\ model with $V=2$ (where SPME exists). (e-h): \modelname\ model with $t_2=1/6$, $V=3.5$ (where no SPME exists). (a) and (e) plot the SIPR of the single-particle eigenstates. (b) and (f) show the energy-resolved average m-SIPR and average MIPR of the many-body eigenstates at half filling, and the shaded area indicates the standard deviation of the corresponding IPR around each energy. In addition, the dashed lines are the CLT predictions given by Eq.~\eqref{Eq:Linear}. 
	(c) and (g) plot the standard deviation of the energy-resolved m-SIPR and MIPR for various system sizes. The calculation is done for $5\%$ of the sampling states around the spectral center, and the black dashed lines are the fittings. 
(d) and (h) plot the DOS at half filling. The range of the horizontal axis in (d) and (h) represents the range of the entire energy spectrum, while the shade range is the energy range used in (b) and (f).
	}
\end{figure*}

\subsection{The energy dependence of MIPR \label{SubSec:B}}

We now use the above revised CLT to argue that the MIPR must be a continuous function of energy, which rules out the high-temperature MBME in a noninteracting fermionic system. 
To start, note that the extent of localization in a single-particle system can be measured by the single-particle IPR (SIPR) defined by
\begin{align}
	I_s=\frac{L\sum_{j}\abs{\psi_j}^4-1}{L-1},
\end{align}
where $\psi_j$ is the wave function on site $j$. 
SIPR vanishes in the thermodynamic limit unless the state is localized.
For a generic noninteracting system, where the many-body eigenstates are all product states, we can introduce the mean SIPR (m-SIPR) of a many-body eigenstate, defined as the mean IPR of its constituent single-particle orbitals. 
Hence, the m-SIPR is an extensive quantity, and therefore we can apply Eq.~\eqref{Eq:CLT} to the bivariate distribution of the energy and the m-SIPR. 
What is more, if the SIPR is an even function of energy, the m-SIPR is independent of the energy. 
However, as long as there is a finite covariance between the SIPR and the energy, the m-SIPR will vary linearly with the energy, regardless of whether an SPME exists in the single-particle model. 
Furthermore, the energy-resolved standard deviation of the m-SIPR (i.e., the standard deviation of the m-SIPR within a small energy window) also scales as $O(1/\sqrt{L})$ because of the CLT. 
Hence, in the thermodynamic limit, almost all states within a small energy window have similar properties. 
Notwithstanding, as MIPR has no additivity, Eq.~\eqref{Eq:CLT} does not apply to MIPR. 
However, the MIPR and the m-SIPR are not irrelevant, as they have the same mean value averaged over all many-body eigenstates~\cite{Vu2022_PRL}. 
The numerical results in Fig.~\ref{Fig:IPR} indeed show the similarity between the MIPR and the m-SIPR. 

As the CLT scales as $O(1/\sqrt L)$, one really needs a large system for the results to be definitive. 
Particularly, we consider the noninteracting \modelname\ model of $L=144$ at half filling in Fig.~\ref{Fig:IPR}, and study two cases, one with an SPME and the other without, as shown in Fig.~\ref{Fig:IPR}(a) and~\ref{Fig:IPR}(e). 
Similar to Fig.~\ref{Fig:CLT}, we randomly choose $10^6$ product states as the sample (with replacement). 
Because of the CLT, almost all the states concentrate within a small energy window compared with the whole spectrum, which can be observed in Fig.~\ref{Fig:IPR}(d) and~\ref{Fig:IPR}(h). 
Within that window, we compute the energy-resolved averaged m-SIPR and MIPR in Fig.~\ref{Fig:IPR}(b) and~\ref{Fig:IPR}(f). 
We have several observations. 
First, the m-SIPR agrees perfectly with Eq.~\eqref{Eq:Linear}. 
Second, not only is the MIPR symmetric relative to the spectral center, but it is also more uniform than the m-SIPR. 
Third, the energy-resolved standard deviation of both m-SIPR and MIPR is very small and follows the $O(1/\sqrt{L})$ scaling as shown in Fig,~\ref{Fig:IPR}(c) and~\ref{Fig:IPR}(g), suggesting that eigenstates within the same energy window have similar localization properties.

From the above observations, we conclude that in a large but finite system, only two degrees of freedom of the single-particle system---the mean value and the variance---play a role in the corresponding noninteracting many-body system. 
Moreover, even the variance has no effect in the thermodynamic limit because of the $O(1/\sqrt{L})$ scaling. 
Additionally, given that the m-SIPR varies smoothly with energy and that a finite high temperature is equivalent to a shift in the center of the spectrum [see Eq.~\eqref{Eq:Temp-Shift}], we conclude that noninteracting many-body systems have neither infinite-temperature MBME nor temperature-dependent MBME (at least at high temperatures). 
The fact that the noninteracting system itself does not manifest an MBME, in spite of perhaps having a single-particle mobility edge, provides a strong hint that there may not be any MBME in the corresponding interacting system.

\subsection{MIPR in a half-filled system \label{SubSec:C}}

In Fig.~\ref{Fig:IPR}, we observed that the plotted MIPR is an even function of energy. 
Actually, this is not a coincidence. 
This feature arises from two crucial properties of a noninteracting many-body system at half-filling. 
The first is that any product state and its complement (under particle-hole operation) must have opposite energies. 
The second is that any product state and its complement have the same MIPR. 

The first property is easy to understand: If a product state occupies half of the orbitals, its complementary state will occupy the rest of the orbitals. 
As a result, the energy of the product state and its complement will have opposite energies. 
To establish the second property, we first need to prove the following identity,
\begin{align}
	\expval{n_i}{\psi}+\expval{n_i}{\psi_c}=1,
\end{align}
where $n_i$ is the particle number and $\ket{\psi_c}$ is the complement of $\ket{\psi}$. 
Suppose that $\ket{\psi}=\prod_{i\in\Omega}f^\dag_i\ket{\text{vac}}$, where $f_j = \sum_{k} U_{jk} c_k$, with $c_k$ being the annihilation operator on site $k$. 
We then have $\ket{\psi_c}=\prod_{i\in\Omega^c}f^\dag_i\ket{\text{vac}}$, where $\Omega_c$ is the complement of $\Omega$. Therefore,
\begin{align}
	\expval{n_i}{\psi}&=\sum_{j,k\in\Omega}U_{ij}^*U_{ik}\expval{f^\dag_jf_k}{\psi}=\sum_{j,k\in\Omega}U_{ij}^*U_{ik}\delta_{jk}\nonumber\\
 &=\sum_{j\in\Omega}\abs{U_{ij}}^2. 
\end{align}
Hence, we obtain
\begin{align*}
	\expval{n_i}{\psi}+\expval{n_i}{\psi_c}=\sum_{j\in\Omega}\abs{U_{ij}}^2+\sum_{j\in\Omega^c}\abs{U_{ij}}^2=1,
\end{align*}
where the last equality follows from the unitarity of $U_{ij}$. Following this, we derive
\begin{align}
	\sum_{i=1}^L(\bar n_i)_c^2
	=&\sum_{i=1}^L(1-\bar n_i)^2\nonumber\\
	=&\sum_{i=1}^L1-2\sum_{i=1}^L\bar n_i+\sum_{i=1}^L\bar{n}_i^2
	=\sum_{i=1}^L\bar n_i^2,
\end{align}
noticing that $\sum_{i=1}^L\bar n_i=L/2$. 
Therefore, these two states have exactly the same MIPR. 
This result may seem counterintuitive because if there are exactly half of the orbitals extended and half localized, then the product state of the extended states has the same MIPR as that of the localized states. 
However, one should not simply infer the property of a product state from a specific basis. 
An example is that the product state of a completely filled band (extended) is the same as that of their Wannier functions (localized). 
Therefore, even if the low-energy orbitals are extended while the high-energy orbitals are localized, the low-energy many-body states are as localized as their high-energy counterparts.

\section{Fermions with short-range interaction \label{Sec:SR} }
\begin{figure}[!]
	\includegraphics[width=\columnwidth]{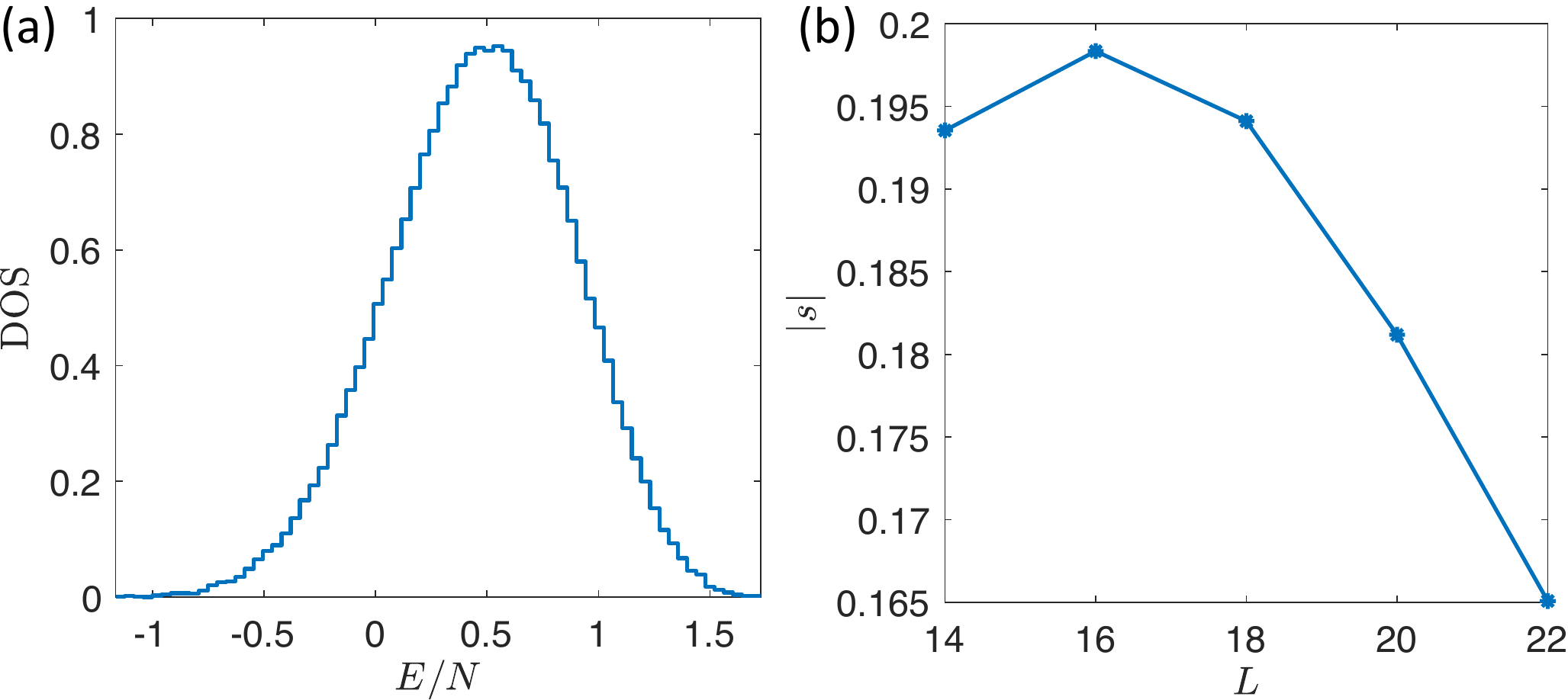}
	\center
	\caption{\label{Fig:Int}
	(a) DOS of the \modelname\ model with $L=18$. (b) Skewness of the spectrum for various system sizes. Here, we take $t_2=1/6$, and $V=U=1$.
	}
\end{figure}

Having understood the properties of noninteracting fermions, a natural question is how the spectrum of a noninteracting many-body fermionic system is modified by the interactions. 
Here we specifically study the \modelname\ model with the nearest-neighbor interaction, $U_{r}=U\delta_{r,1}/2$. 
In Fig.~\ref{Fig:Int}(a), we calculate the DOS in a system of $L=18$ by exact diagonalization. 
It has two distinct features compared with the DOS in the noninteracting spectrum. 
First, the DOS is now an asymmetric function of energy, quantified by a nonzero skewness defined as
\begin{align}
	s=\frac{\expval{(H-\expval{H})^3}}{\left[\expval{H^2}-\expval{H}^2\right]^{\frac32}}.
\end{align}
As shown in Fig.~\ref{Fig:Int}(b), the skewness shows no clear scaling relation for numerically accessible system sizes.
Second, the width of the DOS is much broader than that of the noninteracting results in Fig.~\ref{Fig:IPR}(d) and~\ref{Fig:IPR}(h).

However, we can show that both features are finite-size effects. 
In particular, we prove in Appendix~\ref{SM:IntCLT} that
\begin{align*}
	\lim_{L\to\infty}\expval{\left(\frac{H-\mu'N}{\sqrt{L}}\right)^2}&=\sigma_{\varepsilon}^2\nu(1-\nu)+U^2\nu^2(1-\nu)^2,\nonumber\\
	\lim_{L\to\infty}\expval{\left(\frac{H-\mu'N}{\sqrt{L}}\right)}&=\lim_{L\to\infty}\expval{\left(\frac{H-\mu'N}{\sqrt{L}}\right)^3}=0,
\end{align*}
where $\mu'=\mu_{\varepsilon}+\nu U$, and $\mu_\varepsilon$ is the single-particle mean energy. 
This means that the interacting system also possesses a $1/\sqrt{L}$ scaling, similar to the noninteracting case. 
Hence, in the thermodynamic limit, all states are concentrated within a small energy window centered at $E=\mu'N$, indicating that the infinite-temperature canonical ensemble for an interacting system is equivalent to the microcanonical ensemble at $E=\mu'N$. 
By analogy with the noninteracting case, we conjecture that a finite high temperature is equivalent to shifting the energy of the microcanonical ensemble. 
This indicates that the infinite-temperature MBME, even if it exists, can only occur at $E/N=\mu'+O(1/\sqrt{N})$. 
Moreover, the vanishing skewness of the spectrum suggests that the DOS in a large system would be more symmetric than that in a small system.
Hence, up to the third-order moment, the spectrum of a system with SR interactions is qualitatively similar to that of a noninteracting system in the thermodynamic limit. 

\section{Discussion \label{Sec:Discussion}}

The question of what happens to the single-particle mobility edge separating low-energy localized states from the high-energy extended states in the single-particle spectrum of various well-studied noninteracting quasiperiodic models in the presence of interparticle interactions has remained controversial and open. 
In particular, there is limited numerical evidence in small-system studies that the SPME in the noninteracting system leads to an MBME in the interacting many-body spectrum in the presence of interactions. 
It is also possible that the interactions destroy the SPME, and the interacting many-body spectrum does not manifest any MBME. 
We address this important question with a new approach, where we study the noninteracting many-body spectrum for fermions, establishing that even the many-body noninteracting spectrum does not manifest any MBME, although the corresponding noninteracting single-particle spectrum manifests an SPME. 
This strongly suggests that there is no MBME at all for interacting systems whose noninteracting counterparts have SPMEs. 
We provide further support for this conclusion by explicitly studying interacting one-dimensional quasiperiodic models, which are known to have noninteracting SPMEs in the single-particle spectra.

In this work, we first study the \modelname\ model with LR interaction, extending the previous work on the SR interacting model~\cite{Huang2023} to establish that an infinite-temperature MBME cannot generally emerge from the SPME regardless of the type (LR or SR) of interaction. 
In fact, we demonstrate that the SPME has no qualitative effect on the many-body spectrum, even without any interaction. 
We understand this result by resorting to the CLT, which dictates that the multivariate distribution of the energy and the other extensive quantities commuting with the Hamiltonian follows the multivariate normal distribution. Thus, the CLT obscures and suppresses most details in the single-particle spectrum, including the SPME. 
Moreover, the standard deviation of the spectrum and the energy-resolved standard deviation of the extensive quantities follow the $O(1/\sqrt{L})$ scaling imposed by the CLT. 
As a result, the noninteracting fermionic many-body system generally has neither an infinite-temperature MBME nor a temperature-dependent MBME at high temperatures. 
Furthermore, we also prove that the spectrum of the SR interacting systems still follows the $O(1/\sqrt{L})$ system-size scaling up to the third-order moment, although the numerical results in small systems may not exhibit this feature clearly. 
Hence, the finite-temperature MBME does not imply the infinite-temperature MBME in both interacting and noninteracting systems. 
However, we emphasize that the multivariate distribution of the energy and the MIPR may no longer follow the same $O(1/\sqrt{L})$ scaling in the presence of interactions, which could differentiate the thermal and MBL phases. 

Our work provides a strong hint that there is unlikely to be any MBME in an interacting many-body spectrum, independent of whether the corresponding single-particle noninteracting system has a single-particle mobility edge.  
This conclusion applies to both infinite-temperature MBME and the finite-temperature MBME.

\section*{Acknowledgements}
This work is supported by the Laboratory for Physical Sciences. 
X.L. also acknowledges support from the National Natural Science Foundation of China (Grant~No.~11904305), 
the Research Grants Council of Hong Kong (Grants~No.~CityU~21304720, CityU~11300421, and C7012-21G), 
as well as City University of Hong Kong (Project~No.~9610428). 
K.H. is supported by the Hong Kong PhD Fellowship Scheme. 

\appendix 
\onecolumngrid

\section{Central limit theorem for noninteracting fermions}\label{SM:CLT}
Suppose the noninteracting Hamiltonian $H$ of system size $L$ 
and another observable $Q$ commuting with the Hamiltonian are given by
\begin{align}
	H=\sum_{i=1}^L \varepsilon_in_i, \quad Q=\sum_{i=1}^L q_i n_i.
\end{align}
For the following discussion, we also assume that $\lim_{L\to\infty}\frac1L\sum_{i=1}^L \varepsilon_i^m$ and $\lim_{L\to\infty}\frac1L\sum_{i=1}^L q_i^m$ exist for all positive integer $m$, and that
\begin{align}\label{Eq:Conditions}
	&\lim_{L\to\infty}\frac NL=\nu>0,\quad 
	&&\lim_{L\to\infty}\frac1{\sqrt{L}}\sum_{i=1}^L q_i=0,\quad
	&& \lim_{L\to\infty}\frac1{\sqrt{L}}\sum_{i=1}^L \varepsilon_i=0,\nonumber\\
	&\lim_{L\to\infty}\frac1L\sum_{i=1}^L \varepsilon_i^2=\sigma_\varepsilon^2>0,\quad
 	&&\lim_{L\to\infty}\frac1L\sum_{i=1}^L q_i^2=\sigma_q^2>0,\quad
	&& \lim_{L\to\infty}\frac1L\sum_{i=1}^L\varepsilon_iq_i=0.
\end{align}
Note that $\varepsilon_i$, $q_i$ and $N$ are all functions of $L$ implicitly. Additionally, we also know that $\lim_{L\to\infty}\frac1L\sum_{i=1}^L  \varepsilon_i^mq_i^n$ exists because of the Cauchy inequality. Further, we introduce the following $p$-tuple moment for $x=(x_1,x_2,\cdots,x_p)$
\begin{align}
	f(x)=\sideset{}{'}\sum_{p}\varepsilon_{i_1}^{x_1}\varepsilon_{i_2}^{x_2}\cdots\varepsilon_{i_p}^{x_p},
\end{align}
where $x_j$ are all positive integers, and $\sideset{}{'}\sum_{p}$ denotes $\sum_{\substack{1\leq i_1,i_2,\cdots,i_p\leq L\\ \text{all\ different}}}$.
First, we have
\begin{align}
	f(x)=&f(x_1,x_2,\cdots,x_{p-1})f(x_p)-\sum_{j=1}^{p-1}f(x_1,\cdots,x_i+x_p,\cdots,x_{p-1}),
\end{align}
because
\begin{align}
	\sideset{}{'}\sum_{p-1}\sum_{1\leq i_p\leq L}=\sideset{}{'}\sum_{p}+\sideset{}{'}\sum_{p-1}\sum_{k=1}^{p-1}\delta_{i_p,i_k}.
\end{align}
Next, we are going to prove that $\lim_{L\to\infty}f(x)L^{-\abs{x}/2}$ always exists, where $\abs{x}=\sum_{j=1}^px_j$. For 1-tuples, this statement holds true because of Eq.~\eqref{Eq:Conditions}. If the statement holds for $(p-1)$-tuples, then for $p$-tuple $x$, we have
 \begin{align}\label{Eq:Lemma1}
	\lim_{L\to\infty}\frac{f(x)}{L^{\abs{x}/2}}=\lim_{L\to\infty}\frac{f(x')}{L^{\abs{x'}/2}}\cdot\frac{f(x_p)}{L^{\abs{x_p}/2}}-\lim_{L\to\infty}\sum_{j=1}^{p-1}\frac{f(x_j')}{L^{\abs{x}/2}}.
\end{align}
where $x'=(x_1,x_2,\cdots,x_{p-1})$, and $x_j'=(x_1,\cdots,x_j+x_p,\cdots,x_{p-1})$. Here, we used the fact $\abs{x_j'}=\abs{x}$. As $x'$ and $x_j'$ are all $(p-1)$-tuples, the RHS of Eq.~\eqref{Eq:Lemma1} exists, and therefore the LHS also exists. A direct corollary is that $\lim_{L\to\infty}f(x)L^{-\abs{x}/2}=0$  if $\max\{x\}\geq3$. First, $\lim_{L\to\infty}f(x_1)L^{-\abs{x_1}/2}=0$ if $x\geq3$. If it holds for $(p-1)$-tuples, then for $p$-tuple $x$ with $\max\{x\}\geq3$, we also have Eq.~\eqref{Eq:Lemma1}. Without loss of generality, we suppose $x_p\geq3$, and then $\max\{x_j'\}\geq x_j+x_p\geq3$. Thus, all $(p-1)$-tuples $x_j'$ satisfies $\lim_{L\to\infty}f(x_j')L^{-\abs{x}/2}=0$, and so does the $p$-tuple $x$. It can be proved similarly that $\lim_{L\to\infty}f(x)L^{-\abs{x}/2}=0$ if $|x|$ is odd.

We denote the trace divided by dimension in the fixed particle number subspace as $\expval{\cdot}$. Obviously, for all different $i_1,i_2,\cdots,i_p$, we  have
 \begin{align}
	\expval{p}\equiv\expval{n_{i_1}n_{i_2}\cdots n_{i_p}}=\expval{n_1n_2\cdots n_3}=\frac{N!/(N-p)!}{L!/(L-p)!}.
\end{align}
Whereby, we know $\lim_{L\to\infty}\expval{n_{i_1}n_{i_2}\cdots n_{i_p}}=\nu^p$. Hence, we know that $\lim_{L\to\infty}\frac{\expval{H^{2m+1}}}{L^{m+1/2}}=0$. For $H^{2m}$, we have
 \begin{align}
	H^{2m}=\sum_{p=1}^{2m}\sum_{\substack{x_1\geq\cdots \geq x_p\geq1\\\abs{x}=2m}}c(x)
	\sideset{}{'}\sum_{p}\prod_{j=1}^p\varepsilon_{i_j}^{x_j}\prod_{j=1}^pn_{i_j},
\end{align}
where $c(x)$ is some integer coefficient for $p$-tuple $x$. For convenience, we introduce $x^{(m,p)}=(\underbrace{2,\cdots,2}_{m-p},\underbrace{1,\cdots,1}_{2p})$, and denote $c(x^{(m,p)})$ as $c_{m,p}$. Note that
 \begin{align}
	&H^2\times\sideset{}{'}\sum_{m+p}\prod_{j=1}^{m-p}\varepsilon_{i_j}^2\prod_{j=m-p+1}^{m+p}\varepsilon_{i_j}\prod_{j=1}^{m+p}n_{i_j}\nonumber\\
	=&\sideset{}{'}\sum_{m+p+1}\prod_{j=1}^{m-p+1}\varepsilon_{i_j}^2\prod_{j=m-p+2}^{m+p+1}\varepsilon_{i_j}\prod_{j=1}^{m+p+1}n_{i_j}
     +\sideset{}{'}\sum_{m+p+2}\prod_{j=1}^{m-p}\varepsilon_{i_j}^2\prod_{j=m-p+1}^{m+p+2}\varepsilon_{i_j}\prod_{j=1}^{m+p+2}n_{i_j}\nonumber\\
	&+4p\sideset{}{'}\sum_{m+p+1}\prod_{j=1}^{m-p+1}\varepsilon_{i_j}^2\prod_{j=m-p+2}^{m+p+1}\varepsilon_{i_j}\prod_{j=1}^{m+p+1}n_{i_j}
     +2p(2p-1)\sideset{}{'}\sum_{m+p}\prod_{j=1}^{m-p+2}\varepsilon_{i_j}^2\prod_{j=m-p+3}^{m+p}\varepsilon_{i_j}\prod_{j=1}^{m+p}n_{i_j}+\text{other\ terms}.
\end{align}
Here, ``other terms'' refer to the terms containing $\varepsilon_{i}^3$ or $\varepsilon_{i}^4$. Whereby, we derive the following recursive relation,
\begin{align}
	c_{m+1,p+1}=&c_{m,p}+c_{m,p+1}+4(p+1)c_{m,p+1}+(2p+4)(2p+3)c_{m,p+2},
\end{align}
using which one can prove that $c_{m,p}=\binom{m}{p}\frac{(2m-1)!!}{(2p-1)!!}$ by induction. Thus, we obtain
 \begin{align}
	\lim_{L\to\infty}\frac{\expval{H^{2m}}}{L^m}=&\sum_{p=1}^{2m}\sum_{\substack{x_1\geq\cdots \geq x_p\geq1\\\abs{x}=2m}}c(x)
	\sideset{}{'}\sum_{p}\prod_{j=1}^p\varepsilon_{i_j}^{x_j}\expval{\prod_{j=1}^pn_{i_j}}
	=\sum_{p=1}^{2m}\sum_{\substack{x_1\geq\cdots \geq x_p\geq1\\\abs{x}=2m}}c(x)\nu^p\lim_{L\to\infty}\frac{f(x)}{L^m}\nonumber\\
	=&(2m-1)!!~\nu^{m}\sum_{p=0}^mc_{m,p}\lim_{L\to\infty}\frac{f(x^{(m,p)})}{L^m}.
\end{align}

To calculate $f(x^{(m,p)})$, first notice that
\begin{align}
	f(x^{(m+1,0)})=f(x^{m,0})f(2)-m f(4,2,\cdots,2).
\end{align}
Therefore, we have
\begin{align}
	\lim_{L\to\infty}\frac{f(x^{(m+1,0)})}{L^{m+1}}=\sigma_\varepsilon^2\lim_{L\to\infty}\frac{f(x^{(m,0)})}{L^m}=\sigma_\varepsilon^{2(m+1)}.
\end{align}
Furthermore, because
\begin{align}
	f(x^{(m,p+1)})=f(\underbrace{2,\cdots2}_{m-p-1},\underbrace{1,\cdots,1}_{2p+1})f(1)-(m-p-1)f(3,\underbrace{2,\cdots,2}_{m-p-2},\underbrace{1,\cdots,1}_{2p+1})-(2p+1)f(x^{(m,p)}),
\end{align}
we have
\begin{align}
	\lim_{L\to\infty}\frac{f(x^{(m,p+1)})}{L^{m}}=-(2p+1)\lim_{L\to\infty}\frac{f(x^{(m,p)})}{L^m}=(-1)^p(2p+1)!!~\sigma_\varepsilon^{2m}.
\end{align}
In summary, we obtain
 \begin{align}
	\lim_{L\to\infty}\frac{\expval{H^{2m}}}{L^m}=(2m-1)!!~\nu^{m}(1-\nu)^m\sigma_\varepsilon^{2m}.
\end{align}
Let $\sigma_{\varepsilon}'=\sqrt{\nu(1-\nu)}\sigma_\varepsilon$, and then we have the expected results,
 \begin{align}
	\lim_{L\to\infty}\expval{\left(\frac{H}{\sqrt{L}~\sigma_{\varepsilon}'}\right)^m}=
	\begin{cases}
	0 & m \text{\ is\ odd},\\
	(m-1)!! & m \text{\ is\ even}.
\end{cases}
\end{align}
Similarly,
\begin{align}
	\lim_{L\to\infty}\expval{\left(\frac{Q}{\sqrt{L}~\sigma_{q}'}\right)^m}=
	\begin{cases}
	0 & m \text{\ is\ odd},\\
	(m-1)!! & m \text{\ is\ even},
\end{cases}
\end{align}
where $\sigma_{q}'=\sqrt{\nu(1-\nu)}\sigma_q$.

Finally, we need to calculate $\expval{H^mQ^n}$. 
For the bivariate case, we first introduce the $p$-tuple of rank $2$, given by $z=\begin{pmatrix}
x\\
y
\end{pmatrix}
=\begin{pmatrix}
x_1 &\cdots& x_p\\
y_1 &\cdots& y_p
\end{pmatrix}$, where $x_j, y_i$ are nonnegative integers and $x_j+y_j\geq1$. 
Then the moment of $z$ is given by 
\begin{align}
	f(z)=\sideset{}{'}\sum_{p}\varepsilon_{i_1}^{x_1}q_{i_1}^{y_1}~\varepsilon_{i_2}^{x_2}q_{i_2}^{y_2}\cdots~\varepsilon_{i_p}^{x_p}q_{i_p}^{y_p}.
\end{align}
Similarly, we have
\begin{align}
	f(z)=&f\begin{pmatrix}
x_1 &\cdots& x_{p-1}\\
y_1 &\cdots& y_{p-1}
\end{pmatrix}f\begin{pmatrix}
x_p\\
y_p
\end{pmatrix}-\sum_{j=1}^{p-1}
f\begin{pmatrix}
x_1 &\cdots& x_{j}+x_p &\cdots&x_{p-1}\\
y_1 &\cdots& y_{j}+y_p &\cdots&y_{p-1}
\end{pmatrix},
\end{align}
and therefore $\lim_{L\to\infty}f(z)L^{-\abs{z}/2}=0$ if $\max\{z\}=\max\{x_j+y_j\}\geq3$, where $\abs{z}=\abs{x}+\abs{y}$. Further, we have $\lim_{L\to\infty}f(z)L^{-\abs{z}/2}=0$ if $x_j=y_j=1$ for some $j$. This is obvious for 1-tuple. Providing that it holds for $(p-1)$-tuple, for $p$-tuple $z$ with $x_p=y_p=1$, we have
\begin{align}
	f(z)=&f\begin{pmatrix}
x_1 &\cdots& x_{p-1}\\
y_1 &\cdots& y_{p-1}
\end{pmatrix}f\begin{pmatrix}
1\\
1
\end{pmatrix}-\sum_{j=1}^{p-1}
f\begin{pmatrix}
x_1 &\cdots& x_{j}+1 &\cdots&x_{p-1}\\
y_1 &\cdots& y_{j}+1 &\cdots&y_{p-1}
\end{pmatrix}.
\end{align}
As $x_{j}+y_j+1\geq3$, $\lim_{L\to\infty}f(z)L^{-\abs{z}/2}=0$ also holds true for $p$-tuple. This means that in the calculation of $\lim_{L\to\infty}\expval{H^mQ^n}L^{-(m+n)/2}$ only moment like
\begin{align}
z=\begin{pmatrix}
x & 0\\
0 & y
\end{pmatrix}=
\begin{pmatrix}
\overbrace{2\cdots2}^{p_1} & \overbrace{1\cdots1}^{p_2} & 0\cdots0 & 0\cdots0\\
0\cdots0 & 0\cdots0 & \underbrace{2\cdots2}_{p_3} & \underbrace{1\cdots1}_{p_4}
\end{pmatrix},
\end{align}
where $2p_1+p_2=m$ and $2p_3+p_4=n$.
More importantly, one can easily prove that
\begin{align}
\lim_{L\to\infty}\frac1{L^{\abs{x}/2+\abs{y}/2}}f\begin{pmatrix}
x & 0\\
0 & y
\end{pmatrix}=
\lim_{L\to\infty}\frac1{L^{\abs{x}/2}}f\begin{pmatrix}
x\\
0
\end{pmatrix}
\lim_{L\to\infty}\frac1{L^{\abs{y}/2}}f\begin{pmatrix}
0\\
y
\end{pmatrix}.
\end{align}
This means that the nonvanishing terms in $\lim_{L\to\infty}\expval{H^mQ^n}L^{-(m+n)/2}$ are just all the product of  the nonvanishing terms in $\lim_{L\to\infty}\expval{H^m}L^{-m/2}$ and $\lim_{L\to\infty}\expval{Q^n}L^{-n/2}$. Therefore, we obtain
\begin{align}
	\lim_{L\to\infty}\expval{\left(\frac{H}{\sqrt{L}~\sigma_{\varepsilon}'}\right)^m\left(\frac{Q}{\sqrt{L}~\sigma_{q}'}\right)^n}
	=\lim_{L\to\infty}\expval{\left(\frac{H}{\sqrt{L}~\sigma_{\varepsilon}'}\right)^m}\lim_{L\to\infty}\expval{\left(\frac{Q}{\sqrt{L}~\sigma_{q}'}\right)^n}.
\end{align}
Hence, $H/(\sqrt{L}\sigma_{\varepsilon}')$ and $Q/(\sqrt{L}\sigma_q')$ are two independent normal distribution.
When $\lim_{L\to\infty}\frac1L\sum_{i=1}^L\varepsilon_iq_i\neq0$, $H/(\sqrt{L}\sigma_{\varepsilon}')$ and $Q/(\sqrt{L}\sigma_q')$ form a bivariate normal distribution. This conclusion can be generalized to multivariate cases easily.

\section{Central limit theorem for interacting fermions}\label{SM:IntCLT}
 
The single-particle part of the interacting Hamiltonian can be expressed by
\begin{align}
	H_0=\sum_{i,j}t_{ij}c^\dag_ic_j=\sum_i\varepsilon_if^\dag_if_j,
\end{align}
where $f_j$ is the annihilation operator of the eigenstate of $H_0$. Furthermore, we also need $\varepsilon_i$ to be bounded by some positive number $M$ in addition to Eq.~\eqref{Eq:Conditions}. 
For the interaction part, we only consider the nearest-neighbor (NN) interaction for simplicity, which is given by
\begin{align}
	S=\sum_{i}n_in_{i+1}=\sum_{i}c^\dag_{i+1}c^\dag_{i}c_ic_{i+1}=\sum_{ijkl}S_{ijkl}f^\dag_if^\dag_jf_kf_l.
\end{align}
Hence, the interacting Hamiltonian is given by 
\begin{align}
	H=H_0+US,
\end{align}
where $U$ is the interaction strength.

Let us first calculate $\expval{S}$, $\expval{S^2}$ and $\expval{S^3}$. After some algebra, we obtain
\begin{align}\label{Eq:Moments}
	\expval{S}&=L\expval{2},\quad 
	\expval{S^2}=L\expval{2}+2L\expval{3}+L(L-3)\expval{4},\nonumber\\
	\expval{S^3}&=L\expval{2}+6L\expval{3}+3L(L-1)\expval{4}+6L(L-4)\expval{5}+L(L-5)(L-6)\expval{6}.
\end{align}
As $\lim_{L\to\infty}\expval{S}/(\nu N)=1$, we consider the moments of $\frac{S-\nu N}{\sqrt{L}}$ and $\frac{H-\nu U N}{\sqrt{L}}$ in analogy to the noninteracting CLT. Utilizing Eq.~\eqref{Eq:Moments}, we have
\begin{align}
	\frac1L\expval{(S-\nu N)^2}=&\expval{2}+2\expval{3}-3\expval{4}+L(\expval{4}-2\nu^2\expval{2}+\nu^4),\\
	\frac1L\expval{(S-\nu N)^3}=&\expval{2}+6\expval{3}+3(L-1)\expval{4}+6(L-4)\expval{5}\nonumber\\
    &+(L-5)(L-6)\expval{6}+3\nu^4L^2\expval{2}-\nu^6L^2
	-3\nu^2 L(\expval{2}+2\expval{3}+(L-3)\expval{4})\nonumber\\
	=&L\left[-9(\expval{6}-\nu^2\expval{4})+6(\expval{5}-\nu^2\expval{3})\right.\left.+3(\expval{4}-\nu^2\expval{2})\right]\nonumber\\
	&+L^2\left(\expval{6}-3\nu^2\expval{4}+3\nu^4\expval{2}-\nu^6\right)
	+\text{Constant}.
\end{align}
Using the exact expression of  $\expval{p}$, it is readily to prove
\begin{align}
	\lim_{L\to\infty}\expval{\left(\frac{S-\nu N}{\sqrt{L}}\right)^2}=\nu^2(1-\nu)^2,\quad
	\lim_{L\to\infty}\expval{\left(\frac{S-\nu N}{\sqrt{L}}\right)^3}=0.
\end{align}
Although we just analytically compute the first three moments of $\frac{S-\nu N}{\sqrt{L}}$, the numerical result fits normal distribution very well up to the 5th moment.

Next, we need to calculate $\expval{HS^m}$. First, consider the following operator,
\begin{align}
	O=\sum_{i}\prod_{j=1}^pn_{i+i_j},
\end{align}
where $i_1<i_2<\cdots<i_p$ are some fixed integers. Thus, we can derive
\begin{align}
	\expval{HO}=&\sum_kt_{kk}\expval{n_k\prod_{j=1}^pn_{i+i_j}}\nonumber\\
    =&\sum_i\sum_{k=1}^{p}t_{i+i_k,i+i_k}\expval{\prod_{j=1}^pn_{i+i_j}}
    +\sum_i\sum_{k\neq i+i_1,i+\cdots,i_p}t_{kk}\expval{n_k\prod_{j=1}^pn_{i+i_j}}\nonumber\\
	=&\expval{p}\sum_i\sum_{k=1}^{p}t_{i+i_k,i+i_k}
    +\expval{p+1}\sum_i\sum_i\sum_{k\neq i+i_1,i+\cdots,i_p}t_{kk}\nonumber\\
	=&p\expval{p}\sum_kt_{kk}+(L-p)\expval{p+1}\sum_kt_{kk}
	=L\expval{p}\nu\sum_kt_{kk}=\expval{H}\expval{O},
\end{align}
where $(L-p)\expval{p+1}=(N-p)\expval{p}$ is used. As $S^m$ is just the sum of different $O$'s, we have $\expval{HS^m}=\expval{H}\expval{S^m}$. Consequently, we obtain $\expval{H(S-\nu N)^m}=\expval{H}\expval{(S-\nu N)^m}$.

Another term we need to evaluate is $\expval{H^2(S-\nu N)}$. It is more convenient to compute this term in the eigenstate basis,
\begin{align}
	\expval{H^2(S-\nu N)}&=\expval{H^2S}-\nu N\expval{H^2}=\sideset{}{'}\sum_{i,j}S_{ij}\expval{H^2\tilde n_i\tilde n_j}-\nu N\expval{H^2},
\end{align}
where $S_{ij}=S_{ijji}-S_{ijij}$ and $\tilde n_i=f^\dag_if_i$. Therefore, we have
\begin{align}
	\expval{H^2S}=&\expval{4}\sideset{}{'}\sum_{i,j,k,l}S_{ij}\varepsilon_k\varepsilon_l+\expval{3}\sideset{}{'}\sum_{i,j,k}S_{ij}\varepsilon_k(2\varepsilon_i+2\varepsilon_j+\varepsilon_k)+\expval{2}\sideset{}{'}\sum_{i,j}S_{ij}(\varepsilon_i+\varepsilon_j)^2.
\end{align}
For the third term, we have
\begin{align}
	\abs{\sideset{}{'}\sum_{i,j}S_{ij}(\varepsilon_i+\varepsilon_j)^2}\leq 4M^2\sideset{}{'}\sum_{i,j}S_{ij}=O(L),
\end{align}
where we used $S_{ij}\geq0$ because $S$ is positive semidefinite, and $\sideset{}{'}\sum_{i,j}\abs{S_{ij}}\expval{2}=\expval{S}=L\expval{2}$.
For the second term, we have
\begin{align}
	&\sideset{}{'}\sum_{i,j,k}S_{ij}\varepsilon_k(2\varepsilon_i+2\varepsilon_j+\varepsilon_k)\nonumber\\
	=&\sideset{}{'}\sum_{i,j}S_{ij}\sum_{k}\varepsilon_k^2-\sideset{}{'}\sum_{i,j}S_{ij}(\varepsilon_i^2+\varepsilon_j^2)
	+2\sideset{}{'}\sum_{i,j}S_{ij}(\varepsilon_i+\varepsilon_j)\sum_{k}\varepsilon_k-2\sideset{}{'}\sum_{i,j}S_{ij}(\varepsilon_i+\varepsilon_j)^2\nonumber\\
	=&L\sum_{k}\varepsilon_k^2+O(L)+O(L)\sum_{k}\varepsilon_k.
\end{align}
For the first term, we have
\begin{align}
	&\sideset{}{'}\sum_{i,j,k,l}S_{ij}\varepsilon_k\varepsilon_l\nonumber\\
 =&\sideset{}{'}\sum_{i,j,k}S_{ij}\varepsilon_k\sum_{l}\varepsilon_l-\sideset{}{'}\sum_{i,j,k}S_{ij}\varepsilon_k(\varepsilon_i+\varepsilon_j+\varepsilon_k)\nonumber\\
	=&\sideset{}{'}\sum_{i,j}S_{ij}\left(\sum_k\varepsilon_k\right)^2-\sideset{}{'}\sum_{i,j}S_{ij}(\varepsilon_i+\varepsilon_j)\sum_{l}\varepsilon_l
	-\sideset{}{'}\sum_{i,j}S_{ij}\sum_{k}\varepsilon_k^2+O(L)+O(L)\sum_{k}\varepsilon_k\nonumber\\
	=&L\left(\sum_k\varepsilon_k\right)^2-L\sum_{k}\varepsilon_k^2+O(L)+O(L)\sum_{k}\varepsilon_k.
\end{align}
In conclusion, we obtain
\begin{align}
	\expval{H^2S}=&\expval{4}L\left(\sum_k\varepsilon_k\right)^2+(\expval{3}-\expval{4})L\sum_k\varepsilon_k^2+O(L)+O(L)\sum_{k}\varepsilon_k,\\
	\nu \expval{H^2}=&\expval{2}\left(\sum_k\varepsilon_k\right)^2+(\expval{1}-\expval{2})\sum_k\varepsilon_k^2.
\end{align}
Thus, we have
\begin{align}
	\lim_{L\to\infty}\frac{\expval{H^2(S-\nu N)}}{L^{3/2}}=0.
\end{align}

With all these properties, we can calculate up to the 3rd moment of $\frac{H-\nu U N}{\sqrt{L}}$, which is given by
\begin{align}
	\lim_{L\to\infty}\expval{\left(\frac{H-\nu UN}{\sqrt{L}}\right)^2}&=\sigma_{\varepsilon}^2\nu(1-\nu)+U^2\nu^2(1-\nu)^2,\nonumber\\
	\lim_{L\to\infty}\expval{\left(\frac{H-\nu UN}{\sqrt{L}}\right)}&=\lim_{L\to\infty}\expval{\left(\frac{H-\nu UN}{\sqrt{L}}\right)^3}=0.
\end{align}
Note that ED in small systems finds the prominent asymmetry of the DOS, signified by a finite 3rd moment. However, we prove that the 3rd moment actually goes to zero in the thermodynamic limit, suggesting that the asymmetry of the spectrum is reduced even without an exact symmetry. Notwithstanding, an exactly symmetric DOS dictates that all odd-order moments are zero, so a vanishing 3rd moment is not enough to prove that the spectrum is exactly symmetric.

\section{Spectral Form Factor}

\begin{figure}[!]
\includegraphics[width=0.5\columnwidth]{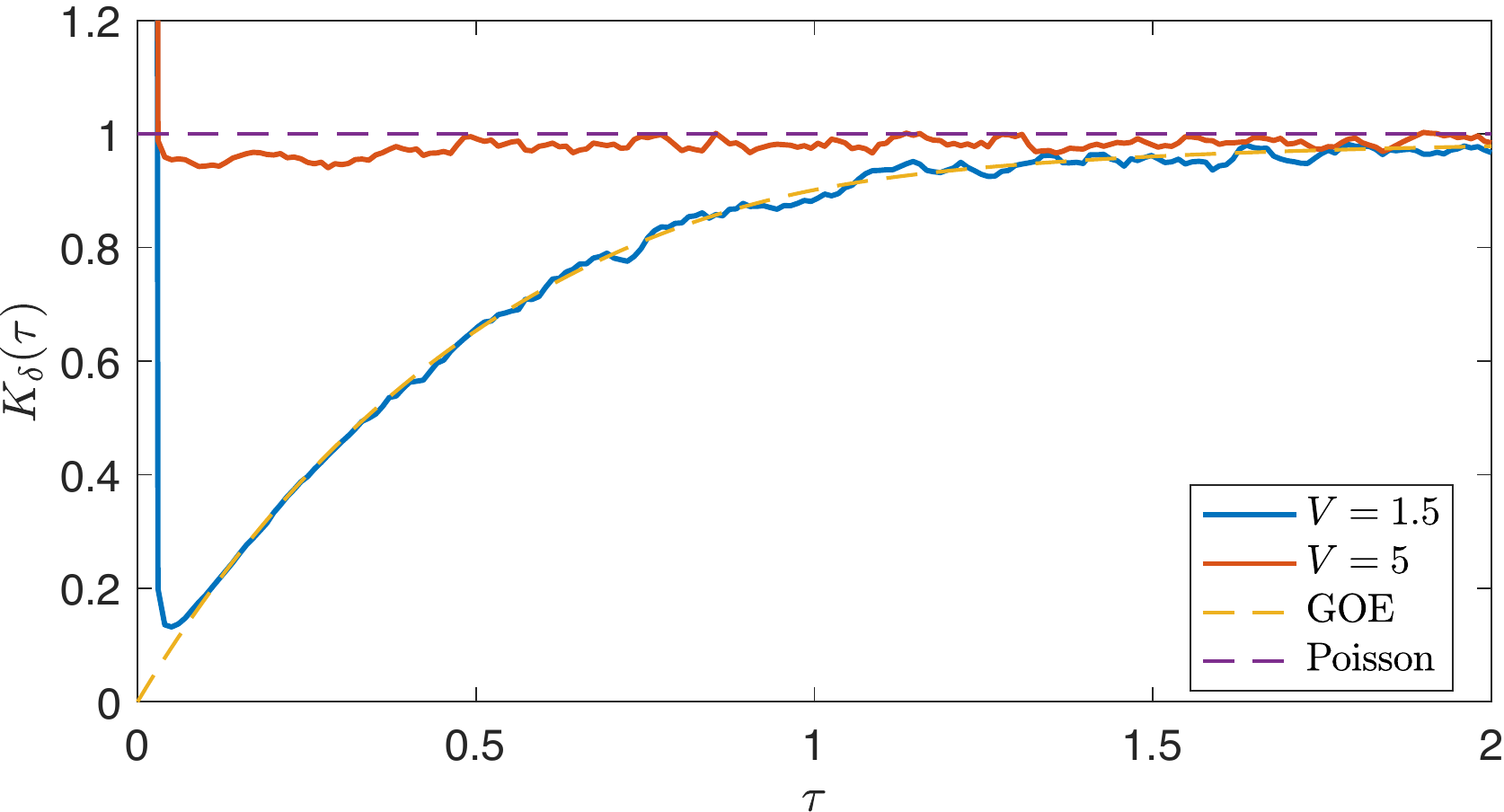}
\caption{\label{Fig:SFF}The time-averaged SFF of the \modelname\ model with LR interaction and $U=1$. The results are calculated in the $L=16$ system and averaged over $100$ random phases. Additionally, we use 2000 eigenvalues in the middle of the spectrum and choose $\delta=1/20$ for the time averaging.
}
\end{figure}

To further investigate the spectral properties of the quasiperiodic systems, we study the spectral form factor (SFF) 
\begin{align}
	K(t)=\frac1D\expval{\abs{\trace{U(t)}}^2}_{\text{ensemble}},
\end{align}
where $U(t)=e^{-itH}$ is the evolution operator, and $\expval{\cdot}_{\text{ensemble}}$ denotes the ensemble average. The SFF is constant $K(t)=1$ for the Poisson distribution, while the analytic form of the SFF of the GOE is
\begin{align}
	K(\tau)&=\begin{cases}
	2\tau-\tau\ln(2\tau+1), & \tau<1\\
	2-\tau\ln\frac{2\tau+1}{2\tau-1}, & \tau\leq1
\end{cases}, 
\end{align}
where $\tau=t/t_H$ and $t_H$ is the Heisenberg time. If the density of state (DOS) is constant $\rho$, then $t_H=2\pi/\rho$. Generally, the SFF contains more information than the mean gap ratio. However, as the SFF is not self-averaging, a considerable number of realizations are necessary to obtain a meaningful result. This is exceptionally complicated in the quasiperiodic systems, because quasiperiodic systems are deterministic and have at most one independent random variable, the random phase. To overcome this difficulty, we utilize the time average technique to reduce the fluctuation. The time-averaged SFF is defined as \cite{PhysRevLett.78.2280}
\begin{align}
	K_\delta(\tau)=\frac1\delta\int_{\tau-\delta/2}^{\tau+\delta/2}K(t)\dd \tau,
\end{align}
where we choose $\delta=1/20$ in our calculation. In Fig.~\ref{Fig:SFF}, we calculate the SFF for the \modelname\ model with LR interaction of length $L=16$. Further, we choose $D=2000$ states in the middle of the spectrum to guarantee an almost constant DOS. Deep in the thermal or MBL phase, Fig.~\ref{Fig:SFF} shows that the SFF agrees excellently with the random matrix theory. Furthermore, we have checked the SFF in the AA, and Anderson models with either LR or SR interaction, and all of them work very well deep in the thermal or MBL phase.

\twocolumngrid

\bibliographystyle{apsrev4-2}
\bibliography{t1t2_Long_v3c}

\end{document}